\documentclass[aip,jcp,reprint,floatfix]{revtex4-1}

\usepackage{amsmath,amssymb,graphicx,color}
\usepackage{dcolumn}
\usepackage{bm}
\usepackage[utf8]{inputenc}
\usepackage[T1]{fontenc}
\usepackage{mathptmx}
\usepackage{etoolbox}
\usepackage{hyperref}
\hypersetup{
    colorlinks = true,
    linkcolor = blue,
    anchorcolor = blue,
    citecolor = blue,
    filecolor = blue,
    urlcolor = blue
}
\usepackage{xcolor}

\makeatletter
\def\@email#1#2{%
 \endgroup
 \patchcmd{\titleblock@produce}
  {\frontmatter@RRAPformat}
  {\frontmatter@RRAPformat{\produce@RRAP{*#1\href{mailto:#2}{#2}}}\frontmatter@RRAPformat}
  {}{}
}%
\makeatother

\begin{document}

\title[Strong-Field Molecular Ionization Beyond The SAE Approximation]{Strong-Field Molecular Ionization Beyond The Single Active Electron Approximation}

\author{J.-N. Vigneau}
\affiliation{Universit\'e Paris-Saclay, CNRS, Institut des Sciences Mol\'eculaires d’Orsay, 91405, Orsay, France.}
\affiliation{Chemistry Department, Université Laval, Québec, QC, Canada.}

\author{T.-T. Nguyen-Dang$^{*}$}
\email{thanh-tung.nguyen-dang@chm.ulaval.ca}
\affiliation{Chemistry Department, Université Laval, Québec, QC, Canada.}

\author{E. Charron}
\affiliation{Universit\'e Paris-Saclay, CNRS, Institut des Sciences Mol\'eculaires d’Orsay, 91405, Orsay, France.}

\author{O. Atabek}
\thanks{Deceased (27-06-2022). Dr. O. Atabek has been the most ardent proponent of the present study of electron correlation effects through an artificial switching-off of the electron repulsion. The interpretation of the results of section III.B of the present article is based on the experience of Dr. Atabek with Zero-Width Resonances (ZWR) (Refs.\,[55] and [61]).  Dr. O. Atabek has been a great mentor for JNV, and a most enthusiastic, insightful  collaborator to us all in this endeavour.}
\affiliation{Universit\'e Paris-Saclay, CNRS, Institut des Sciences Mol\'eculaires d’Orsay, 91405, Orsay, France.}

\date{\today}

\begin{abstract}
The present work explores quantitative limits to the Single-Active Electron (SAE) approximation, often  used to deal with strong-field ionization and subsequent attosecond dynamics. Using a time-dependent multi\-configuration approach, specifically a Time-Dependent Configuration Interaction (TDCI) method, we solve the time-dependent Schrö\-din\-ger equation (TDSE) for the two-electron dihydrogen molecule, with the possibility of tuning at will the electron-electron interaction by an adiabatic switch-on/switch-off function. We focus on signals of the single ionization of H$_2$ under a strong near-infrared (NIR) four-cycle, linearly-polarized laser pulse of varying intensity, and within a vibrationally frozen molecule model. The observables we address are post-pulse total ionization probability profiles as a function of the laser peak intensity. Three values of the internuclear distance $R$ taken as a parameter are considered, $R\!=\!R_{eq}\!=\!1.4$\,a.u, the equilibrium geometry of the molecule, $R\!=\!5.0$\,a.u for an elongated molecule and $R\!=\!10.2$\,a.u for a dissociating molecule. The most striking observation is the non-monotonous behavior of the ionization probability profiles at intermediate elongation distances with an instance of enhanced ionization and one of partial ionization quenching. We give an interpretation of this in terms of a Resonance-Enhanced-Multiphoton Ionization (REMPI) mechanism with interfering overlapping resonances resulting from excited electronic states.\\

\end{abstract}

\maketitle


\section{Introduction}
\label{sec:Intro}

Intense-field dynamics features numerous phe\-no\-me\-na that require radical assumptions or approximations in order to attain  a simple yet comprehensive interpretation  and/or to gain access to affordable quantitative simulations. Strong-field ionization\cite{Ivanov_Spanner_smirnova_JModOPt(2005)} is one of these phenomena. It encompasses a wealth of processes and observable effects, such as High-order Harmonic Generation\cite{PhysRevA.49.2117, Paul-Science.292.1689(2001)} (HHG), Laser-Induced Electron Diffraction\cite{ZUO1996313, Peters-PhysRevA.83.051403(2011), PhysRevA.94.023421} (LIED) or Attosecond Electron Holography\cite{eHolography_NatCommun_9(2018)} for instance. When refering to molecules, these processes define Attosecond Molecular (photo-)Dynamics, and their measurements and simulations  are highly non-trivial\cite{Sansone2012, Martin2017}. Thus, the interpretation and simulations of these processes have often assumed that the properties of the ionization process depend essentially on the interaction of the departing electron with the laser field and with the Coulomb field of the core, possibly screened by an effective, mean field associated with the Coulomb repulsion exerted by the remaining, bound electrons.   This is the celebrated Single-Active Electron (SAE) approximation\cite{PhysRevA.77.063403, PhysRevLett.118.163202, Pegarkov_1999}.

A number of  approaches have been proposed to refine the SAE approximation, making it more quantitative, or to explore its validity. Constructing a model effective potential  for the departing electron in the molecular  ionization is a common thread of these works.  This can  take the form of the introduction of an empirical potential to capture the dynamic (field-induced) multielectron polarization effects\cite{Zhao_Brabec-JMOdOpt_54_981_2007,Zhang_Yuan_Zhao-PhysRev_lett.111.163001(2013),HoangVH_LeVH_ZHaoSF_LeAT-PhysRev.A95.023407(2017)}. Of note is the foundation of this in a rigorous Born-Oppenheimer like separation  between the core and the departing electron, couched within a so-called  Correlated Strong-Field (CSF) ansatz\cite{Zhao_Brabec-JMOdOpt_54_981_2007}.  Else, in an alternative, not less fundamental approach\cite{Ohmura_Ohmura_Kato_Kono-FrontiersPhys.9:677671(2021)}, it was  proposed to use,  as a quantifier of  electron correlation,  the difference in the effective potential derived from the (laser-driven) time-evolution of the system's natural orbitals, as this evolution is  calculated  within  a multiconfiguration time-dependent description (MCTDHF, for Time-Dependent Multiconfiguration Hartree-Fock)  on one hand, and within a time-dependent mean-field description, (TDHF for Time-Dependent   Hartree-Fock) on the other hand.

The present paper explores manifestations of effects that indicate the attainment of quantitative limits to the SAE approximation in H$_2$. Using an approach of the general time-dependent multiconfiguration class\cite{Zanghellini_et_al_JPhysB37(2004), TDMCHF_Nest_Klamroth_Saalfrank_JCP122(2005), Kato_Kono_CPL392(2004), TDMCSCF_TTND_et_al_JCP_2007, Schlegel_Smith_Li_JCP126(2007)} specifically a Time-Dependent Configuration Interaction (TDCI) method\cite{Rohringer_Gordon_Santra_PRA74_2006, TDCI_Krause_Klamroth_Saalfrank_JCP123(2005), Schlegel_Smith_Li_JCP126(2007)}, with a Feshbach partitioning of the many-electron state space\cite{nguyen-dang_multicomponent_2013, doi:10.1063/1.4904102}, we solve the time-dependent Schr\"odinger equation (TDSE) for the two-electron dihydrogen molecule, with the possibility of tuning at will the electron-electron interaction by an adiabatic switch on/off function. It is this possibility to modulate the electron-electron repulsion, during the laser-driven dynamics, which constitutes the distinctive trait of the present work. By doing so, we are probing phenomenological effects of the presence of $V_{ee}$ during the laser-driven ionization dynamics, rather than trying to quantify  electron correlation as  proposed and illustrated nicely
elsewhere\cite{Ohmura_Ohmura_Kato_Kono-FrontiersPhys.9:677671(2021)}.

Our TDCI methodology with Feshbach partitioning\cite{nguyen-dang_multicomponent_2013, doi:10.1063/1.4904102} between neutral bound states and cationic (free) states shares the same philosophy as the Time-Dependent Feshbach Close-Coupling (TDFCC) method of the literature\cite{ Sanz-Vicario2006, Palacios2006,  Palacios_2015, Pauletti2021} and differs from it only in the explicit use of configuration-state functions (CSF) as a basis, as opposed to eigenstates of the field-free two-electron Hamiltonian. We will focus on signals of the single ionization of H$_2$ under a strong NIR (wavelength $\lambda\!=\!700$,\,$750$,\,$790$\,nm) two-cycle FWHM, linearly-polarized laser pulse of varying intensity. To assess the dynamical importance of the electron repulsion $V_{ee}$, we consider three values of the internuclear distance $R$ taken as a parameter, the nuclei being frozen in each geometry in the spirit of the Born-Oppenheimer approximation: $R=R_{eq}=1.4$\,a.u, the equilibrium geometry of the molecule, represents a field-free situation of weak electronic correlation, while $R=5.0$\,a.u and $10.2$\,a.u, denoting respectively an elongated and a dissociating molecule, are situations of increasing field-free electronic correlation.

As expected, in the equilibrium geometry $R=R_{eq}$, tuning off the electron repulsion has little impact on the channel-resolved and total ionization probabilities exhibited as functions of the field-intensity. The total ionization probability vs. intensity profile is typical of a tunnel ionization (TI) process\cite{Ivanov_Spanner_smirnova_JModOPt(2005)} except for the highest range of intensity, for which a sudden increase in the total ionization probability is observed and interpreted as due to the onset of an over-the-barrier ionization (OBI) mechanism\cite{Popov_OBI}. At the elongated geometry $R=5.0$\,a.u, the ionization probability does not exhibit such a sudden increase as one approaches the high-intensity end, i.e. no over-the-barrier enhancement of the ionization is observed. The strong electronic correlation prevailing at this geometry now induces considerable modifications to the dynamics, and the ionization probability varies non-monotonously with $I$, passing through a peak (at $I_{max}$) then a dip (at $I_{min}$), denoting an enhancement and  a quenching of the ionization, in  a moderate intensity range, the value of which depends on the field frequency. A possible explanation is that the ionization is a mixture of a tunnel one (TI) from the (correlated) ground-state and  multiphoton ionizations (MPI) through excited states, accessible from the ground state by multiphoton excitation. The dynamics thus proceeds through a Resonance Enhanced Multiphoton Ionization (REMPI) with interfering overlapping resonances resulting from excited electronic states. This non-monotonous behavior is not observed in the dissociative limit, $R=10.2$\,a.u, the strongest field-free correlation situation, where only subtle differences are observed between correlated and uncorrelated dynamics. This stunning observation is explained by the fact that no multiphoton transition  to the excited states are possible from the correlated ground state, as these transitions between these states become dipole-forbidden in this geometry.

The manuscript is organized as follows: Section \ref{sec:Model} is devoted to the model and to the computational approach chosen for solving the TDSE. The results concerning total ionization profiles are gathered in Section \ref{sec:Results}, with their interpretation and discussions. 
A summary and conclusion are found in Section \ref{sec:Conclusion}.


\section{Model System and Computational Approach}\label{sec:Model}
\subsection{Model System: Orbital Basis and  Configuration-State Functions (CSF)}

The H$_2$ molecule is considered in a body-fixed coordinate system defined such as the $z$ axis coincides with the internuclear axis of the molecule, at varying geometry given by specific values of the internuclear distance $R$. Calculations of the field-free electronic structure of the molecule at the HF-SCF (Hartree-Fock Self-Consistent Field) level, in the $6\hbox{-}31$G$^{**}$ basis and  using the COLUMBUS program suite\cite{COLUMBUS}, yields 10 molecular orbitals. If kept in full, they would in turn generate a basis of $N_{\textrm{CSF}}=55$ two-electron singlet configuration states functions (CSF)\cite{Shepard_CASSCF}. The model we are using  consists of the part of this basis corresponding to the various possible ways to distribute the two electrons in the lowest-lying molecular orbitals of $\sigma$ symmetry, $\sigma_g\ (1s\sigma_g )$ and $\sigma_u\ (2p\sigma_u )$, respectively. In the language of multiconfiguration electronic-structure theories, this corresponds to a CAS(2,2) model of the molecule. These two active orbitals  are the same pair of strongly interacting charge-resonance orbitals of H$_2^+$ that underlie all early works on the dynamics of the dihydrogen molecular ion in an intense laser field\cite{Giusti_Suzor_1995}. Their interactions give rise to several phenomena such as Above-Threshold Dissociation (ATD)\cite{PhysRevLett.64.515}, Bond Softening (BS)\cite{PhysRevA.46.5845}, Vibrational Trapping (VT)\cite{PhysRevLett.68.3869} and Charge-Resonance Enhanced Ionization (CREI)\cite{zuo_charge-resonance-enhanced_1995, PhysRevA.59.2153,PhysRevA.62.031401,PhysRevA.66.043403}. 
Note that 
the interpretation of the  physics of the system using directly these orbitals' properties is possible only near the equilibrium geometry. Near the dissociative limit, the physics is more appropriately discussed in Valence-Bond theoretic language\cite{Shaik_et_al_VB_JCP_157_090901_(2022)}, using Heitler-London orbitals, which asymptotically, (as  $R \longrightarrow \infty$), become atomic orbitals. We will occasionally evoke these in the discussion of the ionization dynamics at elongated geometries and in the dissociative limit.

The  two active orbitals ($\sigma_g,\sigma_u$) give rise also to CSFs that are most fundamental in the discussion of correlation in the smallest two-electron  molecule, H$_2$. Basically, with this active space, the CSF basis generated by these orbitals consists of three states, which will be denoted $|{L}\rangle$, with $L=1,2,3$
\begin{subequations}
\begin{eqnarray}
|{1}\rangle &=&\left|\sigma_{g\uparrow } \ {\sigma}_{g\downarrow} \right| \\
|{2}\rangle &=&\frac{1}{\sqrt{2}}\{\left|\sigma_{g \uparrow} \ {\sigma}_{u \downarrow} \right|-\left|\sigma_{u \uparrow} \ {\sigma}_{g \downarrow} \right|\}  \\
|{3}\rangle &=&\left|\sigma_{u \uparrow} \ {\sigma}_{u \downarrow} \right| 
\end{eqnarray}
\end{subequations}
where $|\xi_1 \ \xi_2|$ designates a Slater determinant constructed out of the orthonormal spin-orbitals $\xi_1,\ \xi_2$, and $\uparrow$ and $\downarrow$ are for spin up or down respectively. These CSF will all be important in the discussion of the electronic excitation dynamics induced by the laser field, insofar as it is polarized along the $z$ direction. In addition, we expect strong ionization to accompany these electronic excitations. To describe the ionization process, the bound orbital active space is augmented by a  reasonably extended set of $n_k$ continuum-type single-electron orbitals taken as plane-waves $|\vec{k}\rangle \,\equiv\, \exp(i\vec{k}.\vec{r})$, pre-orthogonalized with respect to the bound active orbitals, and designated by
\begin{equation}
\label{OPW_def}
|\chi_{\vec k} \rangle \propto (1-\hat q)\, | \vec k \rangle\,,
\end{equation}
where $\hat q$ is the projection operator
\begin{equation}
\hat q =\sum_{i=1}^{2} |\varphi_i \rangle \langle \varphi_i | = |\sigma_g \rangle \langle \sigma_g|+|\sigma_u \rangle \langle \sigma_u|.
\end{equation}
Laser-induced single ionization out of any of the bound-state  CSF $|{L}\rangle, \ L=1,2,3$ will give ionized (singlet) CSFs of the form
\begin{subequations}
\begin{eqnarray}
|{1^+}, \vec k\rangle & \!\!=\!\! &\frac{1}{\sqrt{2}}\left\{\left|\sigma_{g \uparrow} \ \chi_{\vec k \downarrow} \right|-\left|\chi_{\vec k \uparrow}{\sigma}_{g \downarrow} \right|\right\}  \\
|{2^+}, \vec k\rangle & \!\!=\!\! & \frac{1}{\sqrt{2}}\left\{\left|\sigma_{u \uparrow} \ \chi_{\vec k \downarrow} \right|-\left|\chi_{\vec k \uparrow}{\sigma}_{u \downarrow} \right|\right\}
\end{eqnarray}
\end{subequations}
where $\{J^+\!\!=\!\!1^+,2^+\}$ designate the cation CSFs $\{\sigma_g,\sigma_u\}$, respectively, here directly identifiable with the orbitals of H$_{2}^+$. With $\vec k \in \mathbb{R}^3$, these CSFs span two continua of two-electron states corresponding to the ionic channels $1^+$, or $\sigma_g$, and  $2^+$, or $\sigma_u$.

The two-electron wave packet is at all time described by
\begin{equation}
\label{TDCI_wp}
|\Psi(t)\rangle = |\Psi_Q(t)\rangle+|\Psi_P(t)\rangle
\end{equation}
with
\begin{subequations}
\begin{equation}
|\Psi_Q(t)\rangle = \sum_{L=1}^{3}\;c_L(t)\;|L\rangle
\end{equation}
and
\begin{equation}
|\Psi_P(t)\rangle = \int d^3\vec k \sum_{J^+=1^+}^{2^+} \gamma_{J^+,\vec k}(t)\;|{J^+},\vec{k}\rangle,
\end{equation}
\end{subequations}
denoting a Feshbach-Adams partitioning\cite{Feshbach_Ann_Phys:58, lowdin-JMathPhys.3.969(1962), Adams-JChemPhys.45.3422(1966)} of the two-electron state space, into two sub-spaces, the $Q$-subspace (dimension $N_Q=3$) of bound states of the neutral molecule and the $P$-subspace  (dimension $n_k \times N_P$, with $N_P=2$) of singly-ionized states, \emph{i.e.} states of the \{H$_2^+ + e^-$\} system. Since it is  the same partitioning of the many-electron Hilbert space as defined in TDFCC theory\cite{ Sanz-Vicario2006, Palacios2006,  Palacios_2015, Pauletti2021} which refers explicitly to the basis of eigenstates of the field-free two-electron Hamiltonian, the same physical model, (for the same orbital active space), is involved here, although the Hamiltonian matrix is expressed in the CSF basis rather. In particular, the part of the partitioned Hamiltonian corresponding to the interaction between the two subspaces, $\hat H_{QP}$,  is strictly defined by the radiative interaction potential, [see eq.(\ref{H_QP_def_matrix_el}) below].

The configuration-interaction, (CI) coeffi\-cients $c_L(t)$ and $\gamma_{J^{+},\vec k}(t)$ are to be obtained  from their initial values
by solving the electronic Time-Dependent Schr\"odinger Equation (TDSE) describing the electronic system driven by the laser pulse
\begin{equation}
\label{el_TDSE}
i\partial_t|\Psi(t)\rangle=\hat H^{el}_\eta(R,t)|\Psi(t)\rangle\,,
\end{equation}
where the 2-electron Hamiltonian is 
\begin{subequations}
\begin{equation}
\label{hamiltonian_1}
\hat H^{el}_\eta ({R},t)= \sum_{i=1}^{2} \hat h_i({R},t) + \frac{\eta(t)}{r_{12}}\,,
\end{equation}
with
\begin{eqnarray}
\label{hamiltonian_2}
\hat h_i( t) & = & \displaystyle
-\frac{\nabla_i^2}{2} - \frac{1}{|\vec r_i-\vec R/2|} - \frac{1}{|\vec r_i+\vec R/2|} \nonumber \\
& & - \vec r_i \cdot \vec{\mathcal{F} }(t),
\end{eqnarray}
\end{subequations}
containing the spin-conserving electric-dipole interaction (written in the length gauge) between the molecule and the electric component of the laser field $\vec{\mathcal{F}}(t)$. Note the presence, in Eq.\,(\ref{hamiltonian_1}), of the  factor $\eta(t)$ multiplying the electron repulsion potential $V_{ee} = 1/r_{12}$. It allows us to artificially switch the electron interaction on ($\eta =1 $) and off ($\eta =0$) at will, thereby assessing the role (or effect) of electron correlation on the strong-field dynamics. 

\subsection{Many-electron TDSE}

The TDSE, Eq.\,(\ref{el_TDSE}), is solved by the algorithm of Refs. \onlinecite{nguyen-dang_multicomponent_2013, doi:10.1063/1.4904102}.
The time-evolution operator is factorized into a product of a block-diagonal (or intra-subspace) and off-diagonal (inter-subspace)  parts associated with the  Feshbach partitioning defined above. This factorization gives the following structure of the solutions to the partitioned TDSE, here written explicitly, (for $t$ in one of the short time-slices $[t_{n-1}, t_{n}]$  into which the total time evolution is divided), in terms of the vectors $\underline{c}(t), \ \underline{\gamma} (t)$ of CI coefficients $c_L(t)$ and $\gamma_{J^{+},\vec k}(t)$,
\begin{subequations}
\begin{eqnarray}
\underline{c}(t) & = & \mathbb{U}_{QQ}\bigg\{
\mathbb{L}_{0}^{-}[\mathbb{L}_{0}^{+}]^{-1}\;\underline{c}(t_{n-1})\nonumber\\                                 &   & \qquad\quad-i[\mathbb{L}_{0}^{+}]^{-1} \mathbb{H}_{QP}\;
                       \underline{\gamma}(t_{n-1})\bigg\}\label{c_propagated}\\
\underline{\gamma}(t) & = & \mathbb{U}_{PP}\bigg\{\bigg(\mathbb{I}_{P} -
\frac{1}{2}\mathbb{H}_{PQ}[\mathbb{L}_{0}^{+}]^{-1}\mathbb{H}_{QP}\bigg) \underline{\gamma}(t_{n-1})\nonumber\\
& & \qquad\quad +i[\mathbb{L}_{0}^{+}]^{-1}\;\underline{c}(t_{n-1})\bigg\}\,.
\label{gamma_propagated}
\end{eqnarray}
\end{subequations}
In this pair of equations, $\mathbb{H}_{QP}(=\mathbb{H}_{PQ}^\dagger)$ is a $N_Q\times(n_kN_P)$ rectangular matrix, with elements
\begin{subequations}
\begin{equation}\label{H_QP_def_matrix_el}
\left(\mathbb{H}_{QP}\right)_{L,\{J^+,\vec k\}} =
\langle L| \sum_{i} \vec r_i |J^+, \vec k\rangle. \vec E(t_n)\delta t,
\end{equation}
and $\mathbb{L}_{0}^{\pm}$ are $(N_Q\times N_Q)$ matrices defined by
\begin{equation}\label{L0pm_def_matrix}
\mathbb{L}_{0}^{\pm}=  \mathbb{I}_{Q} \pm \frac{1}{4} \mathbb{H}_{QP}\mathbb{H}_{PQ}.
\end{equation}
\end{subequations}
$\mathbb{I}_{Q(P)}$  in the above is the  unit matrix in the $Q(P)$ subspace. The parts containing the matrices $\mathbb{L}_{0}^{\pm}$ and $\mathbb{H}_{QP}$, $\mathbb{H}_{PQ}$ in Eqs.\,(\ref{c_propagated}) and (\ref{gamma_propagated}) correspond to the off-diagonal, inter-subspace propagator, whereas $\mathbb{U}_{QQ}$ and $\mathbb{U}_{PP}$ represent intra-subspace propagators.

In the basis of the  CSFs $|L\rangle$ of the neutral molecule \mbox{M $\equiv$ H$_2$}, $\mathbb{U}_{QQ}$ is given by ($t\in [t_n, t_{n-1}])$
\begin{equation}
\label{U_QQ}
\mathbb{U}_{QQ} = \exp{\{-i \mathbb{H}^{M} (t-t_{n-1})\} },
\end{equation}
while the propagator $\mathbb{U}_{PP}$ of the ionized system \mbox{M$^+$ $\equiv$ \{H$_2^+ + e^-$\}} is factorized approximately into the product of a propagator for the bound cation and one for the ionized electron
\begin{equation}
\label{U_PP}
\mathbb{U}_{PP} = \exp{\{-i \mathbb{H}^{M+} (t-t_{n-1})}\} \otimes u^f(t,t_{n-1}).
\end{equation}
In these equations $\mathbb{H}^{X}$ denotes the (time-dependent) Hamiltonian matrices describing the dynamics of the bound-states of the two-electron molecular system $X$, represented in the bound-CSF basis of that system. For H$_2$, it is a  $(3 \times 3)$ matrix (in the basis of the bound  CSF $|L\rangle$), while for the cation H$_2^+$ system, it is a $(2\times 2)$ matrix in the basis of the  bound CSFs $J^+$ of the cation, \emph{i.e.} of the H$_2^+$ $\sigma_{g(u)}$ orbitals. The one-electron propagator $u^f(t,t_{n-1})$ describes the motion of the ionized electron in the combined action of the Coulomb forces of the ion and that of the field. Given that the strong-field effect is dominant, we use the well-known Volkov form of this propagator, [see Ref. \onlinecite{nguyen-dang_multicomponent_2013} \emph{e.g.}, for details], corresponding to the strong-field approximation (SFA)\cite{Keldysh-JETP.20.1307(1965), Reiss-PhysRevA.22.1786(1980)}.

The initial conditions used are given by
\begin{subequations}
	\begin{eqnarray}
	c_L(0)=c_L^{gs}(R), \label{CI_coeff_initial_bound} \\
	\gamma_{J^{+},\vec k}(0)=0, \ \forall \ J^{+}, \ \forall \ \vec k, \label{CI_coeff_initial_continuum}
	\end{eqnarray}
\end{subequations}
where $c_L^{gs}(R)$ are the CI coefficients of the ground-state of the fully interacting two-electron neutral molecule at internuclear distance $R$, as obtained by diagonalizing the matrix representing the field-free Hamiltonian, i.e. $\hat H^{el}_\eta ({R},t)$ of Eq.\,(\ref{hamiltonian_1}) with $\vec{\mathcal{F}}=0$, and $\eta=1$, in the CSF basis of the Q-subspace. In other words, the initial state is systematically constructed as the ground-state of H$_2$ described at the CISD (Configuration Interaction including Single and Double excitations) level with the 2-orbital active space described above. 

The system then evolves under a two-cycle FWHM  pulse, (for a total width of four cycles), of the form $\vec{\mathcal{F}} = \mathcal{F}(t)\,\hat{e}$, where $\hat{e}$ is a unit vector pointing in the (linear) polarization direction of the field, and where
\begin{equation}
\mathcal{F}(t)=\mathcal{F}_0\,\sin^2\!\left(\frac{\pi t}{\tau}\right)\cos(\omega t)
\end{equation}
with a frequency $\omega=2\pi\,c/\lambda$ corresponding to a wavelength $\lambda$ ranging from $700$ to $790$\,nm. The   pulse duration (total width) is   $\tau$.

The novelty of the numerical simulations presented here lies in the introduction of the parameter $\eta(t)$ multiplying the electron repulsion potential $V_{ee}=1/r_{12}$ in the Hamiltonian of Eq.\,(\ref{hamiltonian_1}), more precisely in the intra-subspace parts $\hat H_{QQ}$, $\hat H_{PP}$ of this Hamiltonian. We will be comparing results of calculations with $\eta(t)=1$ at all $t$, corresponding to the  normal H$_2$ molecule with the two electrons fully interacting with each other, and those of calculations in which $\eta(t)$ is
\begin{equation}
\eta(t)=1-\frac{1}{1+e^{-(12/d)(t-t_c)}},
\end{equation}
with the width parameter $d=75$\,a.u, (about three quarters of the optical cycle),  
chosen sufficiently long to make sure  that no non-adiabatic effects in the time-resolved dynamics are induced by this $V_{ee}$ switch-off process. The center of this sigmo\"id, $t_c=3\pi/2\omega$, is positioned  at   the second maximum of $\mathcal{F}(t)$. With this choice, the $V_{ee}$ switch-off process comes into play   
when the field already acquires important amplitudes, so that the usual assumption of radiative interactions dominating electron correlation  can be meaningfully tested.
The calculations reported in the following use a two-dimensional $k$-grid defining the plane-wave basis. For a field polarized in the same $z$ axis as the molecule alignment, the $k_x$ grid contains $n_x=600$ points with a step-size of $\delta k_x=0.0314$\,a.u while the $k_z$ grid has $n_z=400$ points.

From the propagated time-dependent CI coefficients $c_L(t)$ and $\gamma_{J^{+},\vec k}(t)$, a number of observables can be calculated. The central one in the following  is the final, total ionization probability.

First, let $|E_J\rangle = \sum_L D_{JL} |L\rangle$ describe the composition of the energy eigenstate $|E_J\rangle$ in terms of the CSF, as obtained by diagonalizing $\mathbb{H}_{QQ}$ at the initial time $t=0$. Then the population of that energy eigenstate $|E_J\rangle$ can be calculated from the $c_L(t)$'s by
\begin{equation}\label{P_EJ}
P_{E_J}(t)=\sum_L| D_{LJ}^{*}\ c_L(t)|^2.
\end{equation}
The channel-resolved ionization time-profile, \emph{i.e.} the time-dependent probability of ionization into an ionic channel (ionic CSF) $J^{+}$, is defined  by
\begin{eqnarray}
\label{Ch_resolved_Ionization_prob}
\Gamma_{J^{+}}(t) & \propto & \int d^{3}\vec{k}\;|\gamma_{J^{+},\vec k}(t)|^2 \\
& - & \sum_{i=1}^2\left|\int d^{3}\vec{k}\;\langle\varphi_i |\vec k\rangle\;\gamma_{J^{+},\vec k}(t)\right|^2\nonumber
\end{eqnarray}
where $\varphi_{1/2} = \sigma_{g/u}$.
The total ionization probability is then the sum of all $\Gamma_{J^{+}}(t)$,
\begin{equation}
\label{total_Ionization_prob}
P_{ion}(t)=\sum_{J^{+}}\Gamma_{J^{+}}(t),
\end{equation}
and the proportionality constant implicit in Eq.\,(\ref{Ch_resolved_Ionization_prob}) is such that  $P_{ion}(t)+\sum_{L}P_{E_L}(t) = 1$ at all times. These definitions of    channel-resolved and total ionization probabilities reflect the fact that the CI coefficients $\gamma_{J^{+},\vec k}(t)$ refer to ionic CSF that are  constructed with the non-orthonormal $\{\chi_{\vec k}\}$ continuum orbitals (orthogonalized plane-waves) basis, as defined in Eq.\,(\ref{OPW_def}). However, since asymptotically, $|\chi_{\vec k}\rangle \rightarrow | {\vec k}\rangle$, the channel-resolved photoelectron asymptotic momentum distributions (spectra) are obtained directly from
\begin{equation}
\label{ch_resolved_photoel_distrib}
f_{J^{+}}(\vec k) =| \gamma_{J^{+},\vec k}(t_f)|^2.
\end{equation}


\section{Results and discussions}
\label{sec:Results}

Observables of two types could be used to identify deviations from expected  SAE behaviors. In the following we will focus on scalar observables provided by the populations of various two-electron channels, both in the Q and the P subspaces, (\emph{i.e.} neutral and ionic channels). Signatures of the same deviations from the SAE model found for vectorial observables, in  photoelectron momentum spectra, will be presented and discussed in a forthcoming publication. The results will systematically be reported for three  fixed internuclear distances corresponding to the equilibrium geometry, $R=1.4$\,a.u., a geometry corresponding to an elongated but bound (undissociated) molecule  $R=5.0$\,a.u., and at a sufficiently large distance $R=10.2$\,a.u., where the molecule is almost dissociated.  We wish to emphasize   that, in all calculations, the two-electron multi-channel wave packet is propagated starting from the same initial state of the molecular system, (cf. Eqs.\,(\ref{CI_coeff_initial_bound}) and (\ref{CI_coeff_initial_continuum})), precisely its fully   correlated electronic ground state. While  the radiative coupling of the molecule with the electromagnetic field is being introduced by the laser pulse, we  adiabatically switch off the electron repulsion $V_{ee}$, \emph{i.e.} we modulate the electronic correlation from its natural level at the considered geometry to zero. This allows us to check the validity of SAE  approximation, and to find cases where electron correlation effects are in competition with strong field couplings.

\begin{figure*}[t!]
\includegraphics[trim={0cm 1cm 0cm 0cm},width=\linewidth]{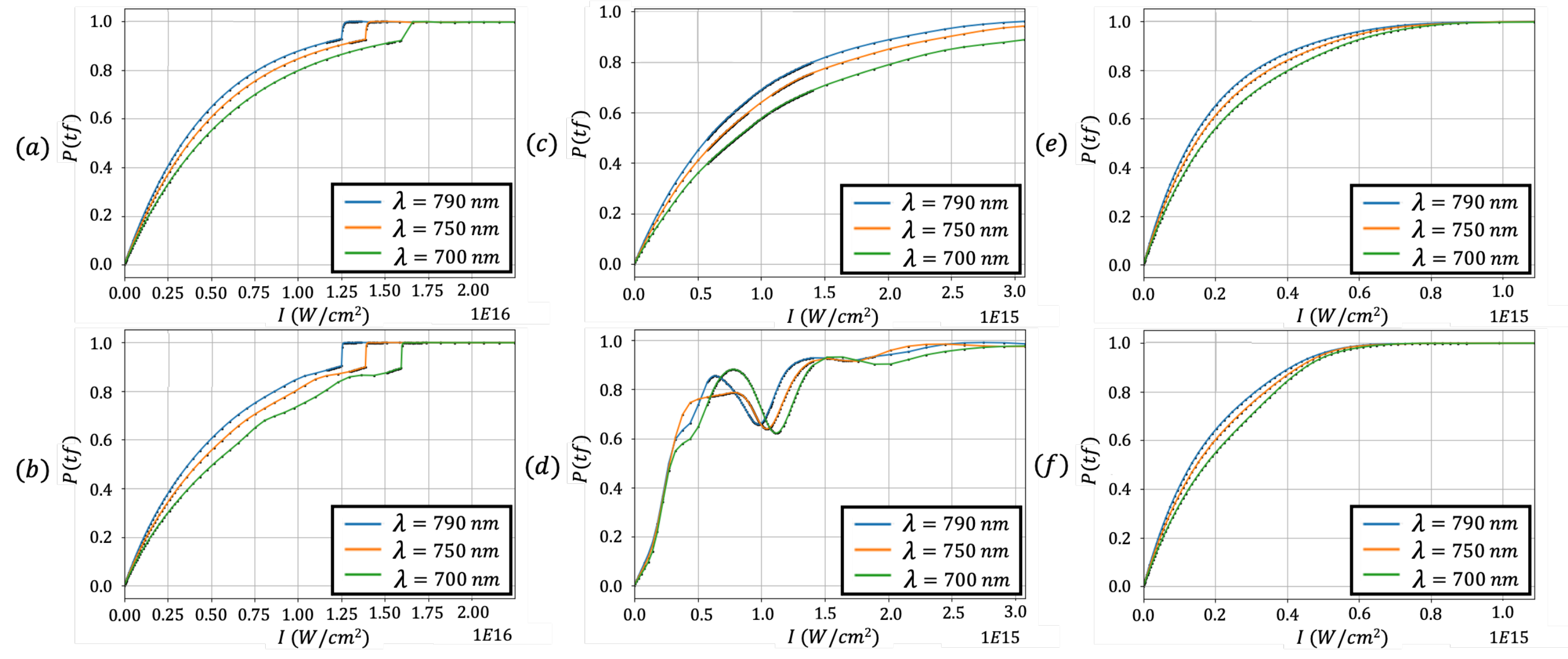}\\
\caption{Total ionization probability profiles given as a function of the laser pulse leading intensity, for three excitation frequencies: $\lambda=790$\,nm blue line, $\lambda=750$\,nm orange line, and $\lambda=700$\,nm green line. The first column [panels (a) and (b)] is for $R=1.4$\,a.u., the second [panels (c) and (d)] for $R=5.0$\,a.u., and the third [panels (e) and (f)] for $R=10.2$\,a.u. The upper row [panels (a), (c) and (e)] corresponds to the adiabatic switch-off of $V_{ee}$, while the lower one [panels (b), (d) and (f)] is for the full dynamical calculation including the electron correlation.}
\label{Total_ionization}
\end{figure*}

The total ionization probability $P_{ion}(t_f)$  is displayed  in Fig.\,\ref{Total_ionization}, as a function of the laser pulse peak intensity, for three values of the carrier-wave frequency, corresponding to wavelengths $\lambda$ within the interval $700 - 790$\,nm. The figure is organized so that the dynamics with the electron interaction $V_{ee}$ switched off is shown in the upper row, while the lower row pertains to the normal fully correlated dynamics. The laser polarization axis $\hat{e}$ is parallel to the internuclear axis.

The first  observation that can be made is that the SAE approximation appears to fail in the case of ionization from H$_2$ in an elongated, non-dissociative geometry, as typified by the case $R=5.0$\,a.u., where the ionization probability shows a non-monotonous behavior for the fully correlated   calculation (lower row), while it (the ionization probability) increases monotonously with the field intensity when the  electron interaction  is switched off. For $R=1.4$\,a.u. (equilibrium geometry) and $R=10.2$\,a.u. (dissociative limit), the ionization profiles in the correlated dynamics are rather smooth,  monotonously growing  with the laser peak intensity.

Let us now examine in detail how these results can be interpreted. A part of the interpretation is based on the Keldysh parameter for the identification of the dominant ionization mechanism underlying the situation in consideration, the other part is based  on the energy positioning of the field-free energy levels with respect to various ionization thresholds. Let us review these two aspects.

The Keldysh parameter is defined as $\gamma=\sqrt{I_p/2U_p}$, where $I_p$ is the ionization potential and $U_p= \mathcal{F}_0^2 / (4\omega^2)$ is the ponderomotive energy, with $\mathcal{F}_0$ the maximum electric-field amplitude. Processes involving $\gamma \ll 1$ are commonly referred to as tunnel ionizations (TI), where the electron tunnels through and  escapes from the barrier resulting from the field-distorted Coulomb nuclear attraction  potential (in a quasi-static picture). In the opposite limit of $\gamma \gg 1$, the electron is thought to be ionized after absorption of several photons, \emph{i.e.} by multiphoton ionization (MPI). Defining a borderline between TI and MPI, for values of the Keldysh parameter in the intermediate range $\gamma \sim 1$, is a challenging issue\cite{Wang_et_al_Opt_Exp(2019)}. The value of $\gamma$ therefore remains an indicator of the dominant mechanism, its increase signaling a change in mechanism from TI to MPI.

Table \ref{Keldysh} collects the values of the Keldysh parameter for a number of important conditions as  defined  by  specific values of the laser peak intensity and ionization potential $I_p$. The value of the latter depends on which initial bound state of the neutral molecule one is considering, in the presence or not of $V_{ee}$. For the three internuclear distances, the table actually considers the $I_p$ of the molecule's ground state only, $|E_1\rangle$ with $V_{ee}$ and $|E_1^0\rangle$ without $V_{ee}$. Note that  $I_p$ changes strongly with $R$, following closely the change in the degree of field-free electronic correlation.

\begin{table}[t!]
\setlength{\tabcolsep}{4.1pt}
\fontsize{9.5pt}{10.25pt}\selectfont
\begin{tabular}{cccccc}
\hline
\hline\\[-0.2cm]
$R$ (au) & State & $I_p$ (au) & $I$ (W/cm$^2$) & $\gamma$ & Mechanism\\
\\[-0.2cm]
\hline\\[-0.2cm]
$1.4$  & $|E_1\rangle$   & 0.60 & $10^{15}$ & 0.37 & TI \\
       &                 &      & $10^{16}$ & 0.12 & OBI \\
       & $|E_1^0\rangle$ & 1.25 & $10^{15}$ & 0.53 & \\
       &                 &      & $10^{16}$ & 0.16 & TI\\
\\[-0.2cm]
\hline\\[-0.2cm]
$5.0$  & $|E_1\rangle$   & 0.49 & $5\times10^{14}$ & 0.47 & MPI \\
       &                 &      & $3\times10^{15}$ & 0.19 &  \\
       & $|E_1^0\rangle$ & 0.75 & $5\times10^{14}$ & 0.59 & MPI \\
       &                 &      & $3\times10^{15}$ & 0.16 & \\
\\[-0.2cm]
\hline\\[-0.2cm]
$10.2$ & $|E_1\rangle$   & 0.52 & $10^{14}$ & 1.09 & MPI \\
       &                 &      & $10^{15}$ & 0.35 & \\
\\[-0.2cm]
\hline
\hline
\end{tabular}
\caption{Keldysh parameter values $\gamma$ for the three different molecular geometries $R=1.4,5.0$ and $10.2$\,a.u. In all cases, the ionization potential  $I_p$ corresponds to the energy of the two-electron molecule's ground state, ($|E_1\rangle$ in the fully correlated case, $|E_1^0\rangle$ in the  uncorrelated case), relative to the threshold of the ionic channel $1^+ \leftrightarrow |\sigma_g, \vec k \rangle$. $I$ are some selected laser peak intensities. The last column is an indication for the possible ionization mechanism (tunnel TI or multiphoton MPI) adopted for the  interpretation.}
\label{Keldysh}
\end{table}

\subsection {Equilibrium geometry}

It is for the equilibrium geometry ($R=1.4$\,a.u.) that the value of the ionization potential is largest, advocating for an MPI reading in a modest intensity range. However, the laser intensities in play here lead to Keldysh parameters as small as $\gamma=0.12$, and we can affirm that the ionization regime is mainly Tunnel Ionization (TI). The monotonous rise of the ionization probability as a function of the field intensity is reminiscent of ionization rate curves deriving from theories such as the ADK model\cite{ADK_JETP(1986)}, and is typical of tunnel ionization. However, one notes a striking probability jump at a wavelength-dependent critical intensity $I_{cr}$ ($I_{cr}=1.25\times 10^{16}$\,W/cm$^2$ for $\lambda=790$\,nm), as seen from Fig.\,\ref{Total_ionization}. We interpret this as marking the onset of an over-the-barrier ionization (OBI) mechanism\cite{Popov_OBI}, a natural high-intensity transformation of TI. In the simple static model, the barrier evoked is that created by the deformation of the nuclear  Coulomb potential, with its two wells centered at $z=\pm R/2$,   distorted by the radiative coupling. To estimate the barrier height, consider this distorted Coulomb potential along the laser polarization ($z$ direction).  Close to $z\simeq R/2$, neglecting the effect of the nucleus  at $-R/2$, such a potential is given by
\begin{equation}
V(z)=-\frac{q_{\textrm{eff}}}{|z-R/2 |} -z\mathcal{F}_0,
\label{Coulomb}
\end{equation}
where $q_{\textrm{eff}}$ represents an effective   nuclear charge, (rather, a nuclear charge number), possibly representing a screening effect of $V_{ee}$ in a simple, semi-empirical mean-field model. This produces a barrier of height $V_{max} = -2 \sqrt{q_{\textrm{eff}} \mathcal{F}_0} - (R/2) \mathcal{F}_0$ positioned at $z_{max}=R/2 + \sqrt{q_{\textrm{eff}}/\mathcal{F}_0}$. For a critical intensity of $1.25\times 10^{16}$\,W/cm$^2$ we get, from the naked Coulomb potential ($q_{\textrm{eff}}=1$) with $R=1.4$\,a.u., a barrier height of $V_{max}=-1.9$\,a.u, located at $z_{max}=1.99$\,a.u. This is to be compared with the ground state energy, $-I_p=-0.6$\,a.u.  with $V_{ee}$ on, \emph{i.e.} for the actual interacting electrons system, and $-I_p=-1.25$\,a.u. without $V_{ee}$, \emph{i.e.} for the a non-interacting two-electron molecule, as indicated in Table \ref{Keldysh}. The height of the barrier being well below the ground state energy, this explains the over-the-barrier ionization mechanism.

\begin{figure}[t!]
\includegraphics[trim={0.5cm 1cm 0cm 0cm},width=\linewidth]{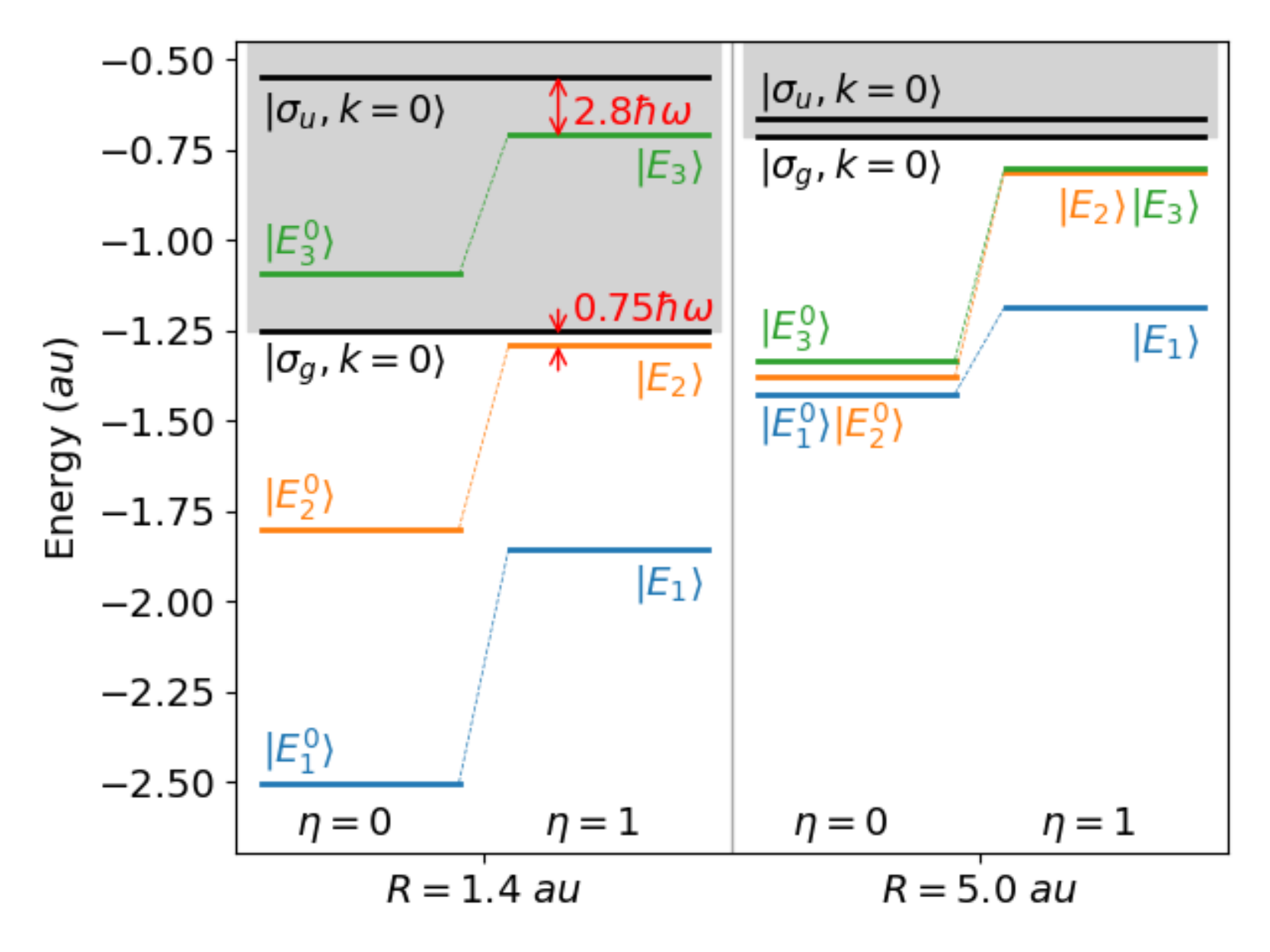}
\caption{Field-free eigenenergies of H$_2$ with (right side labeled as $\eta=1$) and without (left side labeled as $\eta=0$) the electronic interaction potential  $V_{ee}$, at the equilibrium $R=1.4$\,a.u. and extended $R=5.0$\,a.u. geometries. The origin of energies is taken as the second ionization threshold. The gray rectangles correspond to ionization from the first $\sigma_g$ or $\sigma_u$ channels.}
\label{Eigenenergies}
\end{figure}

This information on $I_p$ can also be read  from Fig.\,\ref{Eigenenergies} which shows how the three lowest  energy levels $|E_1\rangle$, $|E_2\rangle$ and $|E_3\rangle$ of the two-electron system  are positioned depending on whether the electrons are considered interacting or not. For the non-interacting electronic system, the ground state is the CSF $|\sigma_g^2\rangle$ with an energy which is twice the $\sigma_g$ orbital energy calculated by diagonalizing the strictly one-electron core Hamiltonian. On the other hand, for the actual interacting two-electrons system, the molecule's energy eigenstates are obtained in a full CI calculation, by diagonalizing the two-electron Hamiltonian matrix in the CSF basis, within the $\{\sigma_g,\sigma_u\}$ active space. This gives
\begin{subequations}
\begin{eqnarray}
|E_1\rangle & = & -0.99\,|\sigma_g^2\rangle + 0.06\,|\sigma_u^2\rangle
\label{CFS1}\\
|E_2\rangle & = & |\sigma_g^1\sigma_u^1\rangle
\label{CFS2}\\
|E_3\rangle & = & 0.06\,|\sigma_g^2\rangle + 0.99\,|\sigma_u^2\rangle
\label{CFS3}
\end{eqnarray}
\end{subequations}
at the equilibrium geometry. The state $|E_1\rangle$ is almost a pure CSF, being dominated by $ |\sigma_g^2\rangle$. It is this fact that confers to the situation at $R=1.4$\,a.u its qualification as a weak-correlation situation. It may come as a surprise that when including the electron repulsion  $V_{ee}=1/r_{12}$ we see a large shift in the ground-state energy (a strong reduction of $I_p$), and yet electron correlation is said to be weak. In this respect, it is important to recall that correlation is to be understood as the correction from the Hartree-Fock limit, where a part of $V_{ee}$ was already included as a mean field (Coulomb and exchange integrals)\cite{Levines_QC_7th_ed}, so that the orbital energies are different from the $V_{ee}=0$ values. Thus this large shift seen in Fig.\,\ref{Eigenenergies} is mainly due to the inclusion of $V_{ee}$ as a mean-field in the HF limit.

Returning to our discussion, with the $I_p$ quoted above,  it can be concluded that, with or without  $V_{ee}$, the ground electronic state  $|E_1 \rangle$ is already well above the barrier for tunnel ionisation. Actually, using the expression of $V_{max}$ given above, a laser field intensity of $4 \times 10^{15}$\,W/cm$^2$ would be sufficient to lower the barrier for the OBI mechanism to operate with respect to ionization from the (correlated) ground state $|E_1\rangle$. The observed critical intensity for the onset of OBI is three times this. There are many reasons for this: (i) The laser electric field being oscillatory, the static model fails when considering the ionization process only at the maximum intensity reached within an optical cycle, and the actual barrier would be, in average, higher than the calculated $V_{max}$. (ii) The Coulomb potential should take into account the presence of the second well located at $z=-R/2$. (iii) The screening effect of the second electron should also be introduced. In addition to these corrections that would affect the barrier height by increasing its value, one should also consider corrections that would affect the energy positioning of  $|E_1 \rangle$. The most important is the radiative Stark effect that would lower the molecular orbital energy level by a non-negligible amount, roughly equal to  $-\mathcal{F}_0R/2 \simeq -0.44$\,a.u.

\begin{figure}[t!]
\centering
\includegraphics[trim={0cm 1cm 0cm 0cm},width=\linewidth ]{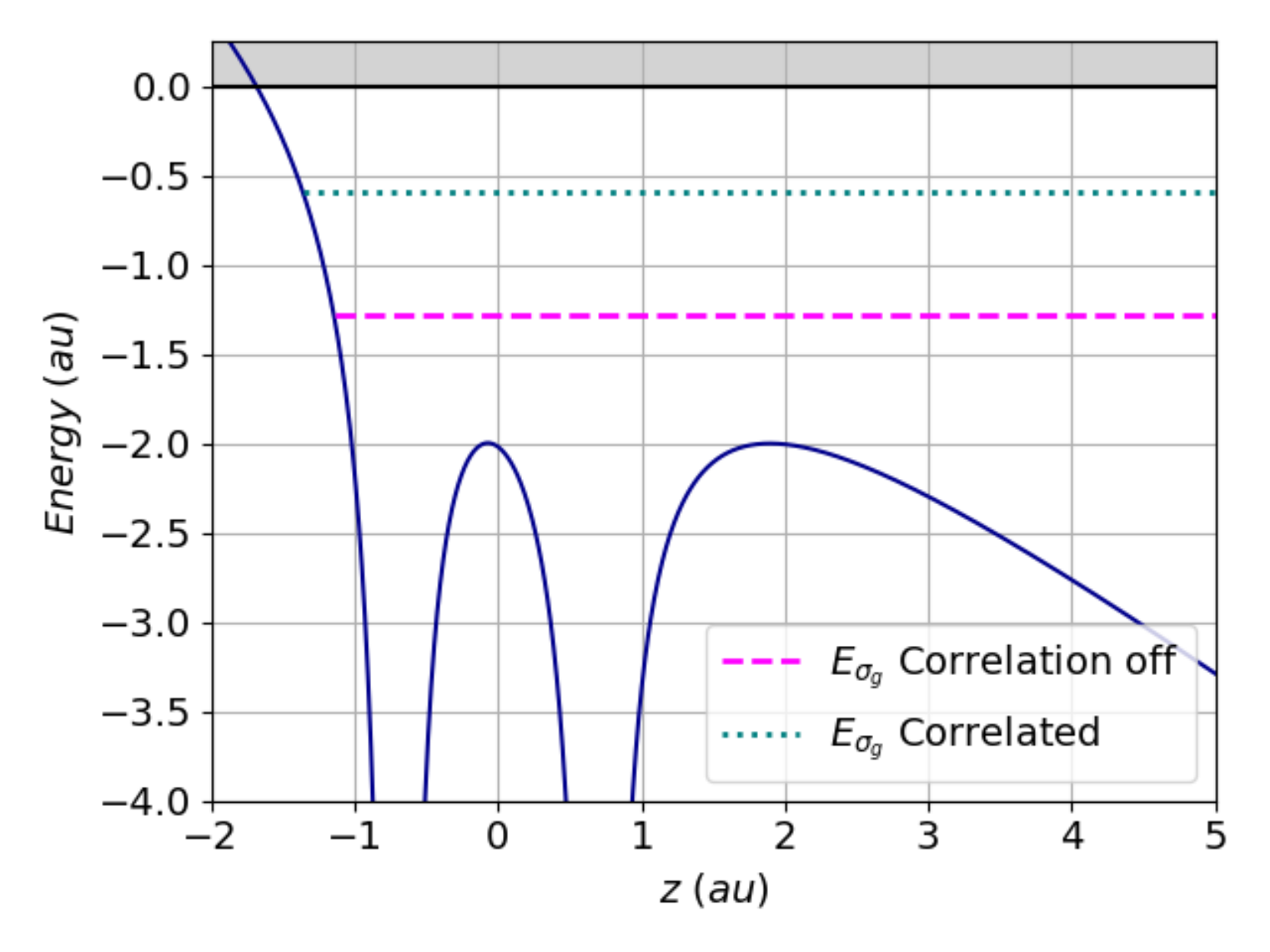}
\caption{Field-distorted Coulomb potential, as defined by Eq.\,(\ref{Coulomb}),   with $R = 1.4$ a.u., $q_{eff} = 1$ and $\mathcal{F}_0$ corresponding to an intensity of
$1.25\times10^{16}$ W/cm$^2$. The ground state level at $-I_p$  is indicated  by the green dotted and the magenta dashed horizontal  lines, respectively for the calculations  with and without electron repulsion. 
}
\label{obti_fig}
\end{figure}

Concerning the two  points (ii) and (iii) made above, we have actually conducted a simple investigation and found that the second nuclear Coulomb attraction center, at $z=-R/2$, with $R=1.4$\,a.u., tends to lower the barrier down, while a screening factor $q_{\textrm{eff}}$ estimated by $q_{\textrm{eff}}^2 = E_1/E_1^0 = 0.6/1.25 \simeq 0.5$, in the spirit of the Quantum Defect Theory\cite{Seaton_1966}, would have an opposite effect, raising the barrier height to a somewhat larger value. Figure\,\ref{obti_fig} shows the distorted Coulomb potential obtained, at $R=1.4$\,a.u., with $q_{\textrm{eff}}=1$ and an intensity of $1.25\times 10^{16}$\,W/cm$^2$. The ground-state energies $E_1$ (interacting electrons) and $E_1^0$ (non-interacting electrons system), placed with respect to the $\sigma_g$ ionization threshold, are indicated by the dotted horizontal line  in green and the dashed line in magenta respectively. Thus, taking $|E_1\rangle$ as the initial state, we will be way above the barrier for TI, at this value of $\mathcal{F}_0$, corresponding to the critical intensity observed for $\lambda=790$\,nm. Adding the Stark shift would give an effective $I_p$ of $\simeq 1$\,a.u., which is still $0.9$\,a.u. above the barrier. The cycle-averaging of the barrier height is thus the only remaining effect that can explain the higher critical intensity found in the calculations.

Another observation to be accounted for is the frequency dependence of the critical intensity marking the onset of OBI.   Figure \ref{Total_ionization} shows that  a higher critical intensity is required when the field frequency increases. From a time-dependent viewpoint, one can argue that, since the potential barrier oscillates at the field frequency, for the lowest frequency ($\lambda=790$\,nm), corresponding to the longest oscillation period $T=2\pi/\omega$, the barrier would stay lowered longer, and a low critical intensity $I_c=1.25\times 10^{16}$\,W/cm$^2$ would be required for OBI to set in, since this intensity is felt on a longer duration. On the other hand, a higher frequency ($\lambda=700$\,nm) corresponds to a faster oscillating barrier, offering less chance for ionization by OBI.

\subsection{Elongated geometries}

Turning now to elongated geometries, we will focus on two typical examples, namely $R=5.0$\,a.u. and $R=10.2$\,a.u. The ionization potential $I_p$ for the ground state decreases progressively as $R$ increases. One notes that the field intensity, $I_{sat}$, where  the ionization probabilities saturates to $\simeq 1$, also decreases. Actually, as can be seen from Fig.\,\ref{Total_ionization}, $I_{sat}$ is, at these elongated geometries, an order of magnitude less than at the equilibrium geometry implying smaller  ponderomotive energies at saturation, and a generally larger Keldysh parameter. As seen in Table \ref{Keldysh}, $\gamma \simeq 0.5$, for an intensity $I=5\times 10^{14}$\,W/cm$^2$ at $R=5.0$\,a.u., and $\gamma \simeq 1.1$, for $I=10^{14}$\,W/cm$^2$ at $R=10.2$\,a.u. Such values correspond to an intermediate regime for the Keldysh parameter, and an MPI interpretation of the processes in consideration is possible. 

\begin{figure}[t!]
\includegraphics[trim={0cm 0cm 0cm 0cm},width=\linewidth]{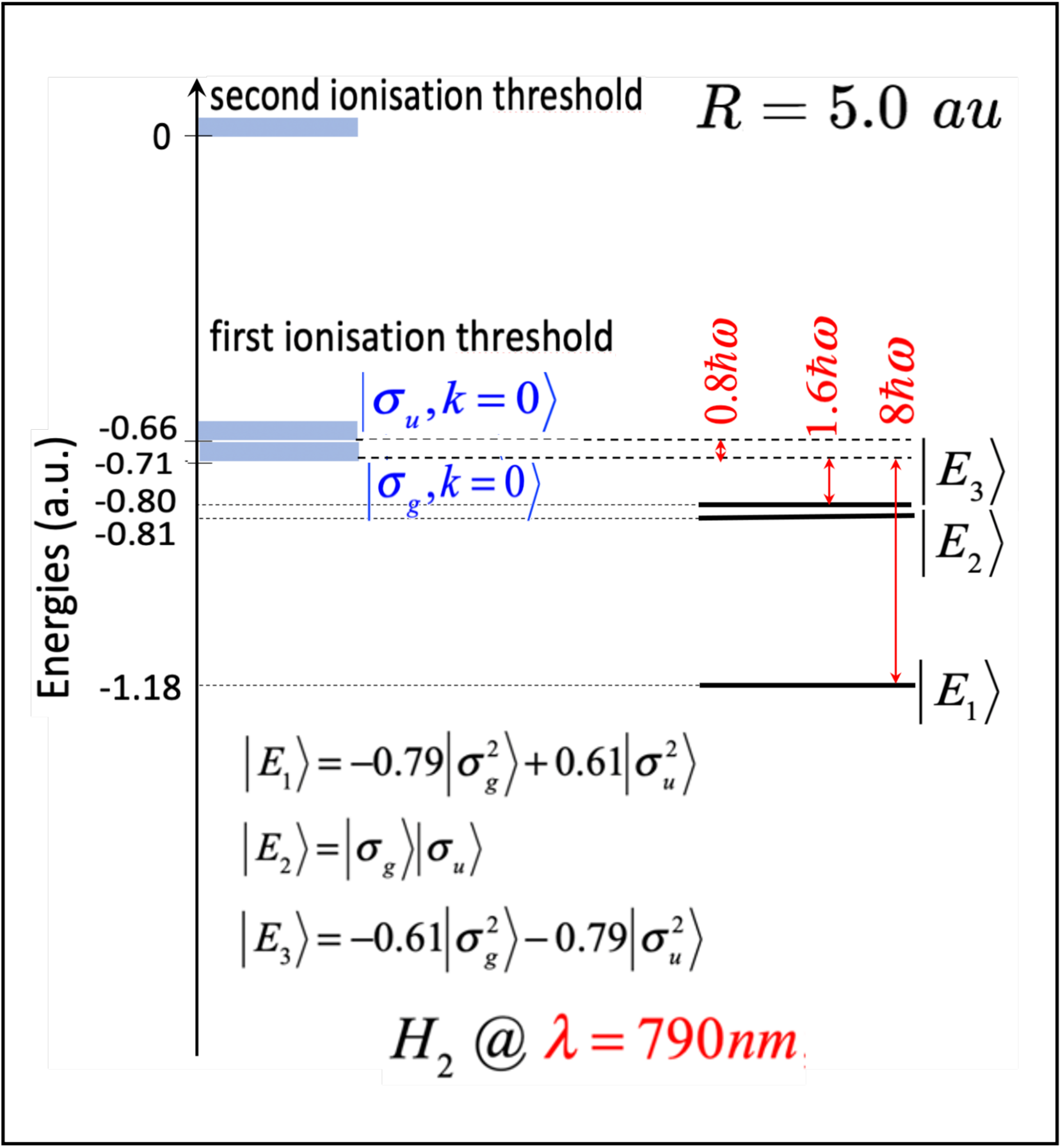}
\caption{Field-free eigenenergies of H$_2$ at $R=5.0$\,au (solid thick black lines) and the ionization thresholds (thick blue rectangles and dashed black lines). The origin of energies is taken as the second ionization threshold. Indicated in thin red arrows are the number of ($\lambda=790$\,nm) photons needed to ionize.}
\label{scheme2}
\end{figure}

The profile of the ionization probability $P_{ion}(t_f)$ as a function of the field intensity exhibits, at $R=5.0$\,a.u, an intriguing behavior around   $I=5\times 10^{14}$\,W/cm$^2$, in the fully correlated dynamics (cf. Fig.\,\ref{Total_ionization}, panel (d)). The total ionization probability follows at first a smooth rising curve at low intensity, to deviate from this curve at around $3\times10^{14}$\,W/cm$^2$, exhibiting from then on a rather strong oscillatory pattern (localized around $0.5-1.5\times10^{15}$\,W/cm$^2$). This behavior is not observed in the uncorrelated ($V_{ee}$ switched off) case (shown on panel (c) of Fig.\,\ref{Total_ionization}), for which the ionization regime is TI at high intensity. It is not observed either in the dissociative limit, $R=10.2$\,a.u. It  appears to be correlated with the degree of excitations to the bound excited states, $|E_j\rangle, j=2,3$, which  in the correlated system are found much closer to the ionization thresholds than in the  non-interacting electrons system, (cf. Fig.\,\ref{scheme2}). We will see that these excitations are extinguished completely at $R=10.2$\,a.u.

The energy diagram for the lowest three  energy eigenstates is displayed in Fig.\,\ref{scheme2}, together with the first $|\sigma_g, k=0\rangle$ and second $|\sigma_u, k=0\rangle$ ionization thresholds. The full CI  calculation in field-free condition gives the composition of these energy eigenstates at $R=5.0$\,a.u as
\begin{subequations}
\begin{eqnarray}
|E_1\rangle & = & -0.79\,|\sigma_g^2\rangle + 0.61\,|\sigma_u^2\rangle
\label{CFS11}\\
|E_2\rangle & = & |\sigma_g^1\sigma_u^1\rangle
\label{CFS21}\\
|E_3\rangle & = & -0.61\,|\sigma_g^2\rangle - 0.79\,|\sigma_u^2\rangle
\label{CFS31}
\end{eqnarray}
\end{subequations}
Contrary to what is observed in the case of the equilibrium geometry (Eqs.\,(\ref{CFS1}) to (\ref{CFS3})), we have an important configuration mixing, a signature of a strong electron correlation. The expansion coefficients in Eq.\,(\ref{CFS11}) denote an uneven (asymmetric) distribution of population among the configurations $\sigma_g^2$ and $\sigma_u^2$ in $|E_1 \rangle$, the $\sigma_g^2$ configuration still dominating at $64\%$.

At this range  of $R$, the symmetry allowed \mbox{$\sigma_g \leftrightarrow \sigma_u$} transition dipole moments in $z-$linear polarization, acquires important values, (it is well known that it increases linearly with $R$, and this is confirmed by the preparatory \textit{ab initio} calculations), and  the ground state $|E_1\rangle$ is directly and  strongly  coupled to the first excited state $|E_2\rangle$. The excitation of this state, which is the configuration $\sigma_g^1\sigma_u^1$, either from the $|\sigma_g^2\rangle$ or $|\sigma_u^2\rangle$ components of the ground state $|E_1\rangle$, instantly debalances this initial state in its CI content, amounting to populating (suddenly) the other excited state $|E_3\rangle$ as well, by the same multiphoton process that has prepared $|E_2\rangle$.

\begin{figure}[t!]
\centering
\includegraphics[trim={0cm 1cm 0cm 0cm},width=\linewidth]{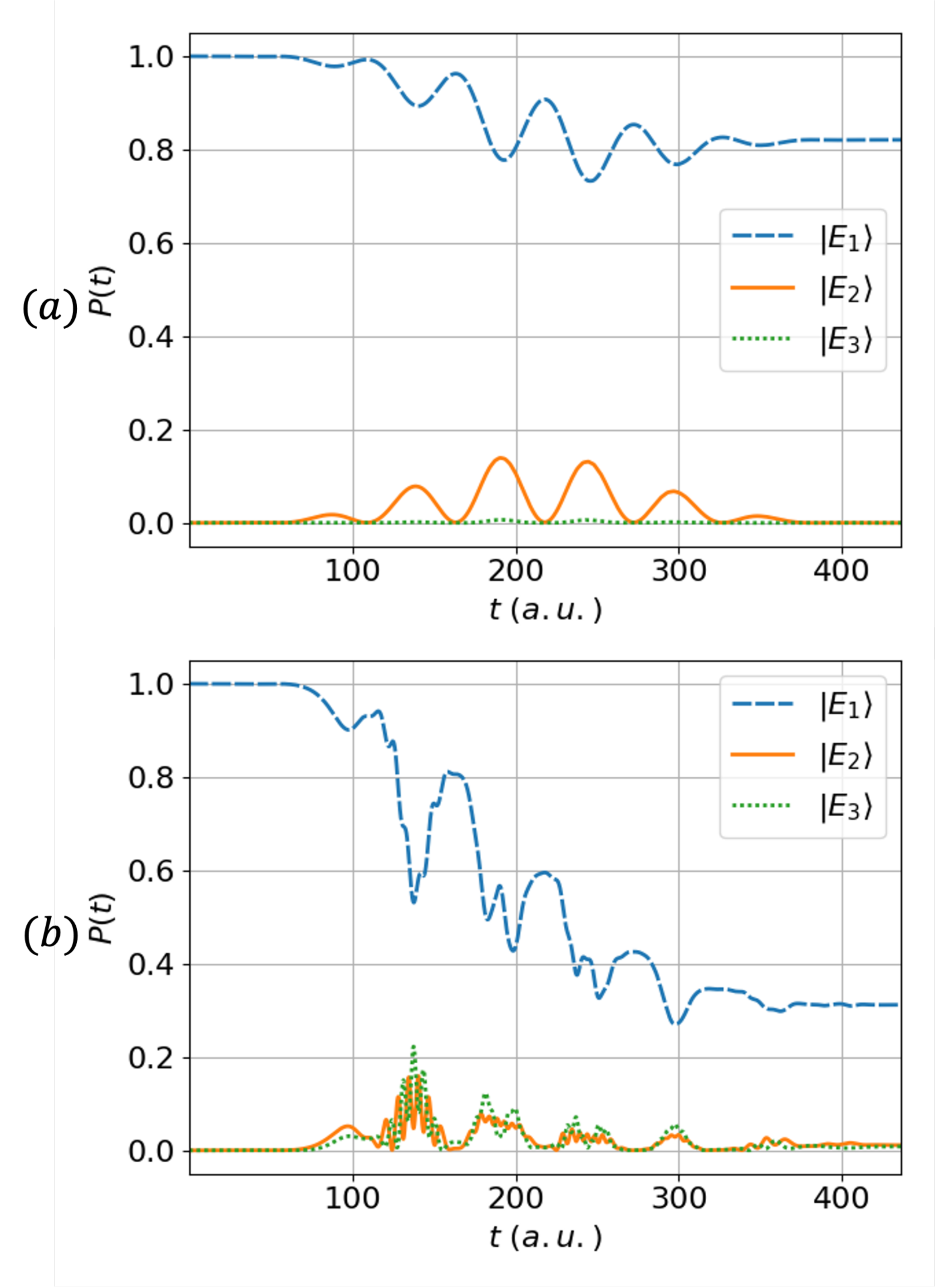}
\caption{Time evolution, during the $I=10^{15}$\,W/cm$^2$, $\lambda=790$\,nm laser pulse, of the populations of the energy eigenstates at (a) $R=1.4$\,a.u., and (b) $R=5.0$\,a.u.}
\label{fig_P_E_k_R_5_eq} 
\end{figure}

This dynamics of laser-driven excitation among the bound states is well illustrated by Fig.\,\ref{fig_P_E_k_R_5_eq}, panel (b), which shows the time evolution, during the $I=10^{15}$\,W/cm$^2$, $\lambda=790$\,nm laser pulse, of the populations of the energy eigenstates at $R=5.0$\,a.u, to be compared to    the same bound-states dynamics for the $R=1.4$\,a.u in panel (a). In the equilibrium geometry, the two bound states that have  appreciable populations during the pulse are $|E_1\rangle$ and $|E_2\rangle$. Their populations  oscillate in phase with the field oscillations, with only the ground state population exhibiting a decay in the mean, due to ionization. In contrast, at $R=5.0$\,a.u, all three eigenstates are populated appreciably, with the populations of $|E_2\rangle$ and $|E_3\rangle$ of equal magnitude and tracking each other, throughout the pulse, and all three decay in the mean. Superimposed on this mean decay curve denoting the ionization out of  the  designated eigenstate, one notes population oscillations or rather beating at two frequencies, the lowest being $\omega_{32}= E_3-E_2$, the highest $\omega_{31}=E_3-E_1$. The fact that the bound-state dynamics is strongly driven by the two states $|E_1\rangle$ and $|E_3\rangle$ is further seen in the dynamics of the bound CSFs. 

\begin{figure}[t!]
    \includegraphics[trim={0cm 0cm 0cm 0cm},width=\linewidth]{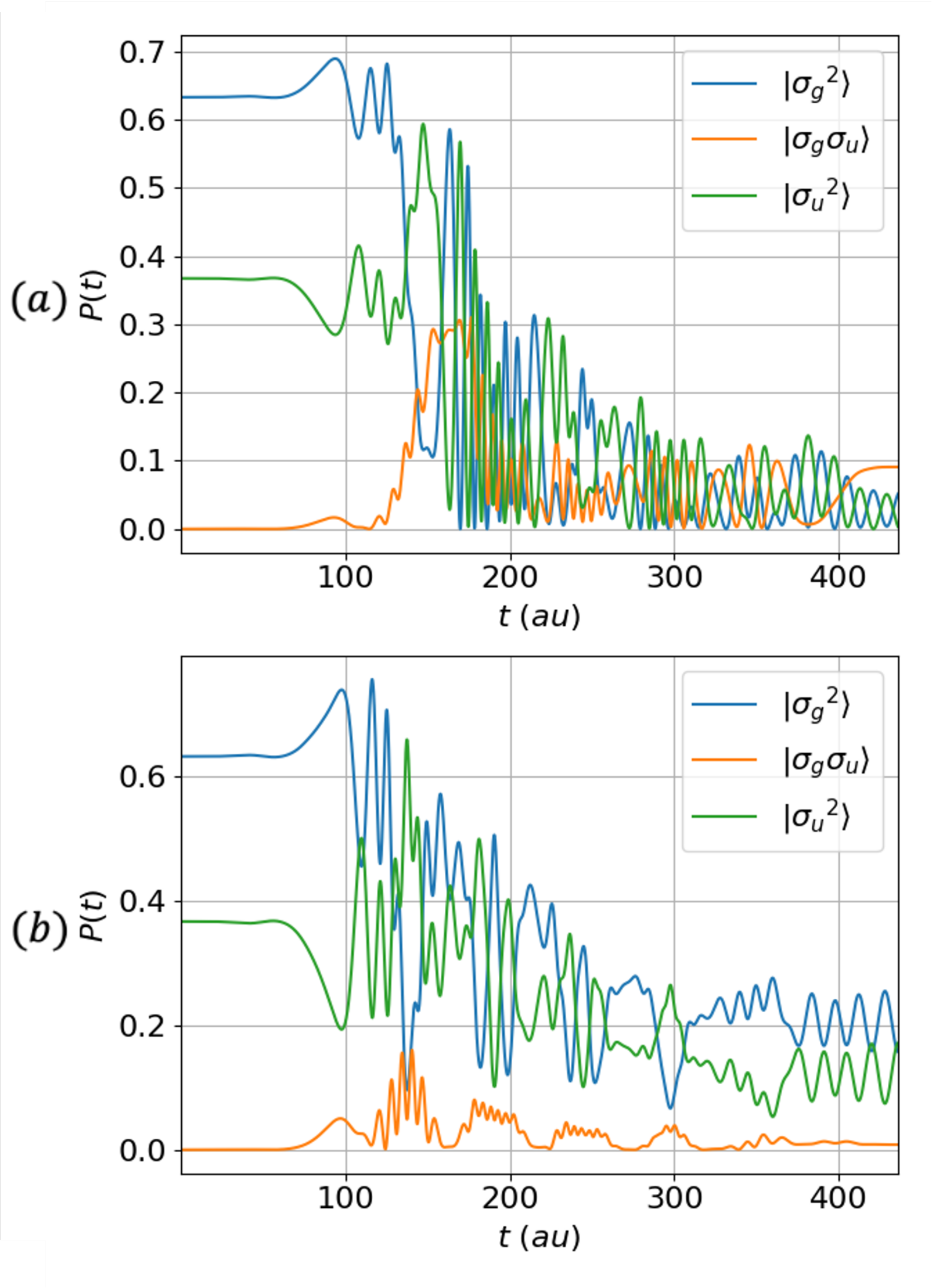}
    \caption{Time evolution of the populations of  CSFs $|\sigma_g^2\rangle$, $|\sigma_g \sigma_u\rangle$ and $|\sigma_u^2\rangle$ of $H_2$ at $R = 5.0$\,a.u during   a $\lambda = 790$\,nm laser pulse, at (a)   $I = 6.25\times10^{14}$ W/cm$^2$, where ionization is enhanced, and (b)   $I = 1.00\times10^{15}$ W/cm$^2$, where ionization is quenched.}
    \label{csf}
\end{figure}

Fig.\,\ref{csf} shows, for the $\lambda=790$\,nm case, the populations $|c_I  (t)|^2, (I=1,2,3)$ of these CSFs as a function of time for $I=6.25\times 10^{14}$\,W/cm$^2$, and $I= 10^{15}$\,W/cm$^2$, corresponding to a maximum and minimum in the ionization probability profile of Fig.\,\ref{Total_ionization}, panel (d). 
In the two cases, the populations of the two CSFs $\sigma_g^2$ and $\sigma_u^2$ that compose $|E_1\rangle$ and $|E_3\rangle$  exhibit very strong  oscillations in phase opposition at the two frequencies identified above. In all, these results show clearly how strongly and coherently the two excited states $|E_2\rangle$ and $|E_3\rangle$ are accessed from the ground state, giving a dynamics that is strongly dependent on the excitation energies $E_j-E_1, \ j=2,3$ (6-7 photons) and the  difference between them, $E_3-E_2$.
We will come back  to this Fig.\,\ref{csf} later, to discuss how comparing  panels (a) and (b) can explain the different ionization yields at these two intensities.

The interpretation we have in mind for the observed  oscillatory behavior in the ionization probability profile, in the case $R=5.0$\,a.u, is based on a Fano\,\cite{Fano_PhysRev.124.1866} type picture, involving two interfering routes. While the initial population on  $|E_1\rangle$ can be ionized (by TI) directly, it can also be first transferred on $|E_2\rangle$ through a seven-photon absorption process. An additional photon is sufficient to ionize the molecule from $|E_2\rangle$  to the $|\sigma_g,\chi_k\rangle$ channel. With two photons, the ionization would ionize the system to the $|\sigma_u,\chi_k\rangle$ channel instead. This route is  a Resonance-Enhanced-Multiphoton-Ionization (REMPI)\,\cite{REMPI_doi:10.1063/1.438436}. Else, the state $|E_3\rangle$ can also be populated strongly once $|E_2\rangle$ is, (recall Fig.\,\ref{fig_P_E_k_R_5_eq}). From $|E_3\rangle$, which is composed of the same CSF as $|E_1\rangle$, we may also have a single-photon (two-photon) ionization to the $\sigma_g (\sigma_u)$ ionic channel. We are thus facing a situation with two field-dressed resonances ($E_2$ and $E_3$), close in energy and decaying into the same ionization continua $|\sigma_g(\sigma_u),\chi_k\rangle$. These overlapping resonances  
  have almost the same energy (real part of the complex resonance energy), but different widths. We can expect that one of these resonances have a much larger width than the other one\footnote{One can think of the same effect in this electron (ionization) dynamics as the one which gives rise to Zero-Width Resonances (ZRW) in the context of  vibrational, i.e laser-driven dissociation dynamics\cite{CL_PRA77_043413}.}. Referring to the discussion to be found in the following paragraph, we can identify  the  resonance with the largest width as an ionic doorway state, while the stabilizing resonance  is a covalent state. Now, these resonance states with energies and widths defined within the Floquet representation\cite{CL_PRA71_023403, CL_PRA77_043413}, would be transported adiabatically in time during the time-development of the pulse envelope. 
Through non-Abelian  Berry  phases associated with this adiabatic transport\cite{Berry_ProcRSocA392_45}, these overlapping resonances may thus interfere with each other to give an intensity-dependent ionization rate: some intensities would lead to the ionic doorway state (more ionizing) being more populated at pulse-end, some others to the preponderance of the covalent (less ionizing) state, producing the St\"uckelberg-type oscillations\cite{Stuckelberg_1932}   in the ionization profiles of Fig.\,\ref{Total_ionization}, panel (d). We have assumed adiabatic transport of the resonances under the  pulse. Given the relatively short duration of the pulse, we expect non-adiabatic effects to be non-negligible both on the rising and the descending sides of the pulse. These could further impact  the resonance population dynamics,   again in a strongly intensity-dependent manner.

It is also important to note that in the absence of electron correlation this specific resonance overlapping mechanism cannot happen. This is clear from Fig.\,\ref{Eigenenergies}, where the lowest three eigenstates,  $|E_1^0\rangle$,  $|E_2^0\rangle$,  $|E_3^0\rangle$, are seen well separated, and lie much lower in energy,  beside the fact that in compositions, they are pure CSFs, completely different from the fully correlated states $|E_1\rangle$, $|E_3\rangle$ which denote strong mixing of CSFs.

An alternative explanation, in a time-dependent semi-classical approach, can also be attempted using the electron trajectory view of the three-step rescattering mechanism\cite{PhysRevLett.71.1994}. At such a geometry, the intensity and frequency dependent ionized electron quiver radius is roughly comparable to the  size of the elongated molecule. In other words, contrary to the situation where the ionized electron feels an almost point-like molecule, (the case at $R=1.4$\,a.u), or one where it rather sees two almost separated, also point-like atoms (the case at $R=10.2$\,a.u), in the case of an  intermediate elongated geometry, the electron trajectory would somehow remain within a ``cage of the molecule", not being able to leave it. This could explain  at least an ionization quenching, as observed at certain values of the field intensity. Actually that extended ``cage of the molecule" exists only in so far as it corresponds to the covalent elongated molecule, \emph{i.e.} the covalent part of the initial correlated wavefunction at $R= 5.0$\,a.u. That wavefunction has an ionic part (which would tend asymptotically to a H$^-$ + H$^+$ dissociation state), for which the system again reduces to a point-like two-electron H$^-$ anion accompanied by a bare proton. That ionic part is known to be a doorway state for enhanced ionization (CREI) of this two-electron molecule\cite{PhysRevA.62.031401,PhysRevA.66.043403}. It could be increased or decreased by a time-dependent admixture of the excited $|E_2\rangle, \ |E_3\rangle$ states. The dominance of the   ionic or covalent component,  during the time-dependent dynamics at certain intensity could give rise to an enhancement or a quenching of the ionization probability, as observed. 

This can be clearly seen in Fig.\,\ref{csf}. We first note that only the CSFs $|1\rangle \equiv |\sigma_g^2\rangle$ and  $|3\rangle\equiv |\sigma_u^2\rangle$ can give an asymptotically  covalent   configuration $1s_A 1s_B$, where $1s_{A(B)}$ denotes a  Heitler-London, or rather Coulson-Fischer orbital of Valence-Bond theory\cite{Shaik_et_al_VB_JCP_157_090901_(2022)} that becomes the $1s$ atomic orbital  centered on proton H$_A$ or H$_B$.  A purely covalent state 
is attained when the oscillating populations of these two CSFs are equal, i.e. when the time-profiles of their populations cross each other. It is clear, from perusal of Fig.\,\ref{csf}, that at $I=10^{15}$\,W/cm$^2$, this state acquires a higher population in average, (giving a stabilization with respect to ionization), than at $I=6.25\times 10^{14}$\,W/cm$^2$. In both panels, the coherent fast oscillations of the populations of CSFs $|1\rangle$ and  $|3\rangle$ would give roughly a small  average contribution, to the (asymptotically) ionic configuration ($1s_A^2 +1s_B^2$) at both intensities. This contribution is to be added to the contribution of CSF $|2\rangle \equiv |\sigma_g \sigma_u\rangle$ whose population   clearly rises to larger average values during the pulse at $I=6.25\times 10^{14}$\,W/cm$^2$, the intensity of a maximum 
in the ionization probability profile of Fig.\,\ref{Total_ionization}, panel (d), than at  $I=10^{15}$\,W/cm$^2$, where $P_{ion}(t_f)$ is at a minimum.

To summarize, just as for the interpretation in terms of resonance interferences discussed above, 
the strong electron correlation plays a central role in    this alternative interpretation. This electron correlation is already manifest through the strong CI mixing in the initial state, perturbed by field-driven excitations to $|E_2\rangle$.

\subsection{Dissociation limit}

\begin{figure}[t!]
\centering
\includegraphics[trim={0cm 0cm 0cm 0cm},width=\linewidth]{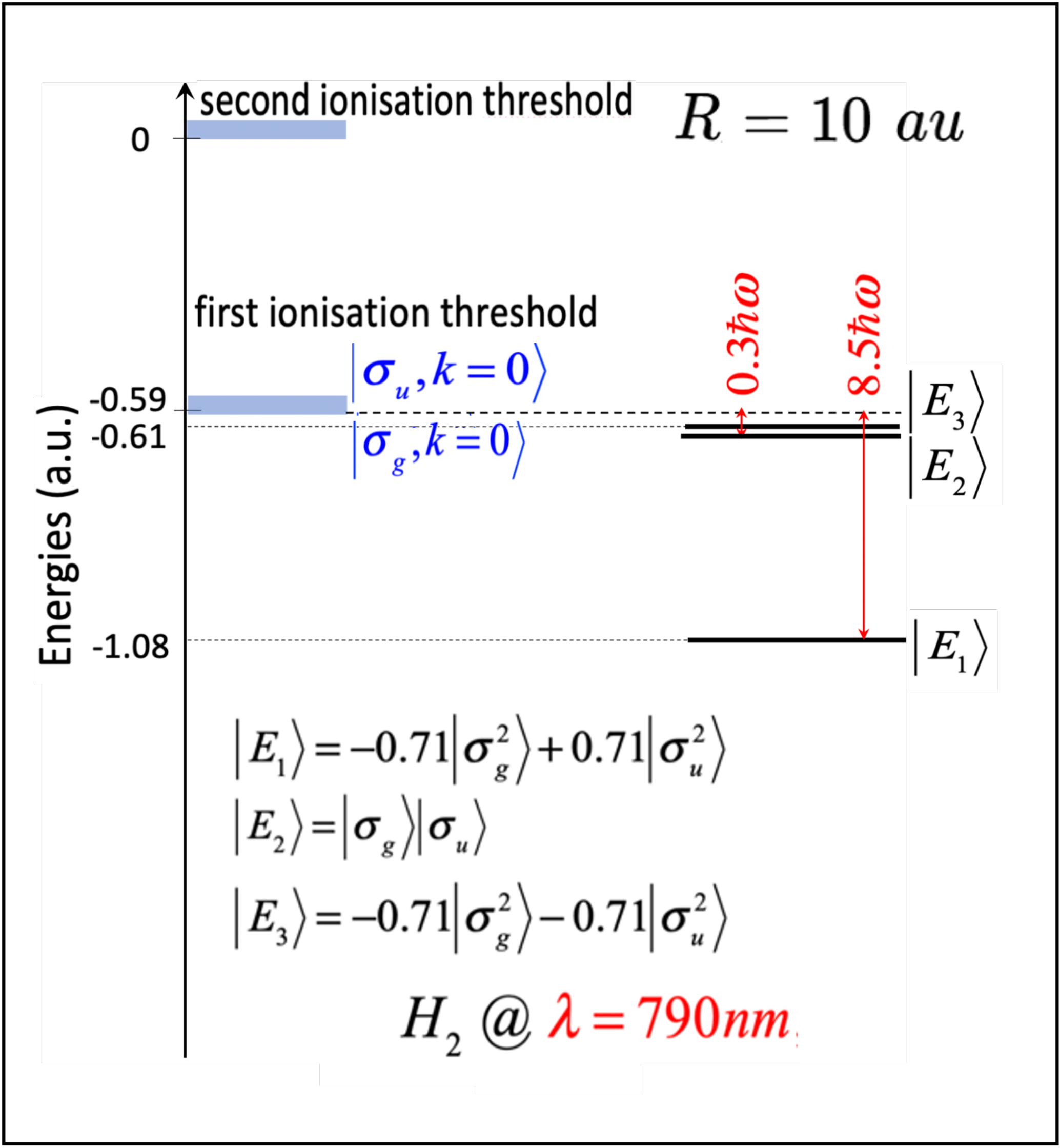}
\caption{Same as for figure (\ref{scheme2}) but for the internuclear distance $R=10.2$\,a.u.}
\label{scheme3}
\end{figure}

For the largest internuclear distance  $R=10.2$\,a.u, we are practically in the dissociative limit. The electron correlation is strongest, as testified by the CI composition of the three two-electron eigenstates
\begin{subequations}
\begin{eqnarray}
 |E_1\rangle & = & -0.71\,|\sigma_g^2\rangle + 0.71\,|\sigma_u^2\rangle \label{CFS12}\\
 |E_2\rangle & = & |\sigma_g^1\sigma_u^1\rangle
 \label{CFS22}\\
 |E_3\rangle & = & -0.71\,|\sigma_g^2\rangle - 0.71\,|\sigma_u^2\rangle
 \label{CFS32}
 \end{eqnarray}
\end{subequations}
featuring equal (even) but anti symmetric con\-tri\-bu\-tions of the CSFs $|\sigma_g^2\rangle$ and $|\sigma_u^2\rangle$ to the ground-state. The configuration mixing is maximal, denoting the highest electron correlation effect. The energy diagram of Fig.\,\ref{scheme3} suggests that an even stronger resonances overlap situation should occur, as the levels $|E_2\rangle$ and $|E_3\rangle$ are closer than for $R=5.0$\,a.u. Yet the ionization probability profiles are smooth curves, monotonously increasing with the field intensity. No  oscillations due to resonances overlap is observed.  This can be understood by the fact that the transition dipole moments $\mu_{12}$ from  $|E_1\rangle$ to  $|E_2\rangle$ vanishes identically, due to the anti-symmetric contributions of the $|\sigma_g^2\rangle$ and $|\sigma_u^2\rangle$ CSFs to $|E_1\rangle$. Indeed, using Condon-Slater rules in Ref. \onlinecite{Levines_QC_7th_ed}, p.321, it can easily be shown that 
\begin{equation}
\langle \sigma_g^1 \sigma_u^1|\mu  |\sigma_g^2\rangle= \langle \sigma_g^1 \sigma_u^1|\mu  |\sigma_u^2\rangle
\end{equation}
so that
\begin{equation}
\mu_{12} \propto \big[\langle\sigma_g^1\sigma_u^1|\mu|\sigma_g^2\rangle-\langle\sigma_g^1\sigma_u^1|\mu |\sigma_u^2\rangle)\big] = 0.
 \end{equation}
No transition to the excited states is thus possible, and the ionization is a direct 9-photon ionization or tunnel ionization from the ground-state only. The interference process we are referring to for the intermediate internuclear distance $R=5.0$\,a.u. no longer holds, and the ionization profile smoothly increases without any oscillation, saturating for an intensity close to $7\times10^{14}$\,W/cm$^2$. 

A number of remarks ought to be made at this point. First, the apparent lack of excitation at the dissociative limit in the fully correlated calculation is due to the fact that, within the simplified model considered, as resulting from the choice of the minimal orbital active space, the ground state is that of a H atom located either at $\pm R/2$, and that in this model, H has only one orbital, $1s$. In a complete model, the parallel laser field can give transition to higher lying states of the atom, (e.g. $2p_z$), corresponding to the dissociative limit of higher energy MOs. These are simply not included (intentionally) in the active space of the present model. Without excitation to $|E_2 \rangle$ or $|E_3 \rangle$ possible from the $|E_1 \rangle$,  ionization at $R=10.2$\,a.u. can occur only from the ground state, and corresponds to tunnel ionization or 9-photon ionization from the $1s$ atomic orbital.  With $V_{ee}$ switched off, we have exactly the same tunnel ionization from the same initial state as in the fully correlated case, hence the transferability of the ionization profiles from one case to another, as seen on the last column of Fig.\,\ref{Total_ionization}. The effect of electron correlation to produce an energy level scheme among which laser-induced excitations correspond to interfering REMPI processes and/or overlapping and interacting multiphoton ATI (Above-Threshold Ionization)\cite{EBERLY1991331} resonances, giving rise to non-monotonous ionization vs. intensity profiles, is only seen in elongated geometries not too close to the dissociative limit. This non-monotonous ionization probability profile deviates strongly from a single electron TI profile and constitutes a clear signature of a non-SAE behavior.

\subsection{Experimental considerations}

How to experimentally observe  the non trivial behavior of the ionization probabilities at an elongated (but non-dissociative geometry) as a function of the field intensity? Our primary purpose here has been to understand how strong electronic correlation could modulate the strong-field ionization dynamics. The simplicity of the model used here, with the nuclear motion  frozen and the electronic excitation dynamics reduced to its essential elements, helps in this respect. We have, in particular, shown that strong St\"uckelberg type of oscillations are present at $R= 5.0$\,a.u. The same oscillations in the intensity profile of the ionization probability are found,  although with different amplitudes, for  other values of  $R$ (in the range $4-6$\,a.u.) that have also been considered in our calculations (not showed here). This electron-correlation driven behavior of the ionization dynamics is thus not   an artifact of the case $R=5.0$\,a.u that was discussed in length in the text. That it does not appear in the strongest correlation case at $R=10.2$\,a.u is due to the extinction of the coupling between the correlated ground and excited states, an interference effect, as it is conditioned by the opposite phases of the $|\sigma_g^2 \rangle$ and $|\sigma_u^2 \rangle$ components of the ground state. As the   electron-correlation driven  behavior of the ionization probability typified by the $R=5.0$\,a.u case appears to set in as soon as one departs from $R_{eq}$ in the higher $R$ range, it would be encountered to some extend during the vibrational motion of the molecule. Thus the non-monotonous variation of $P_{ion}(t_f)$ with $I$, though expected to be weaker  within the support of the vibrational ground state of the molecule, and somewhat   washed out by the averaging over the vibrational motion, may still be observable. This remains to be assessed by more detailed calculations, including vibrational motions.

We can also imagine that the range of large values of $R$, where the oscillations in $P_{ion}(t_f)$ \emph{vs.} $I$ are strong, could be accessed by a Raman vibrational excitation of the molecule, using a first laser pulse operating in the XUV (pump pulse). With a proper time-delay, the NIR laser as considered here, (probe pulse), can then interrogate the ionization dynamics at an elongated geometry. It is  to be noted that a vibrational excitation  exceeding the $v=10$ level of H$_2$ would be necessary to expect an average value of $R$ reaching  $5.0$\,a.u. and beyond.  

Still another way to prepare the molecule in an elongated geometry is to exploit long-range and long-lived scattering Feshbach resonances resulting from a bound state embedded in the translational energy continuum, associated with the laser-controlled collision between a pair of free H atoms\cite{CL_PRA77_043413}. We have recently studied such laser bound H$_2^+$ molecules (LBM) in the context of laser cooling\cite{PhysRevA.101.063406}. It was  shown that such a laser bound quasi-stable hydrogen molecular system LBM (as opposite to the usual chemically bound molecule CBM), can be obtained using a THz laser with a wavelength of $\lambda=25\,\mu$m and an intensity of about $3$\,GW/cm$^2$. The associated wavefunction has a spatial extension which can grow up to an average internuclear distance $\langle R\rangle \simeq 13$\,a.u.

These schemes for probing the molecule's ionization dynamics at an elongated but non-dissociative geometry, in a strong correlation regime, can become   more interesting if  we can record in coincidence  photoelectron spectra taken at a sub-femtosecond time scale  and providing a snapshot of the molecule undergoing dissociation. Such channel-resolved photoelectron or LIED spectra, if emanating strictly from a single molecular orbital, as  implied by the SAE\cite{Peters-PhysRevA.83.051403(2011), puthumpallyjoseph:tel-01301505}, would  exhibit equally spaced interference fringes from which the internuclear distance $R$ can be inferred. In addition, this fringe pattern comes with a definite, clear nodal structure that is a signature of the molecular orbital from which the photoelectron is ionized\cite{doi:10.1080/00268976.2017.1290837, doi:10.1080/00268976.2017.1317858, puthumpallyjoseph:tel-01301505}. The part of our research dealing with these photoelectron spectra, within the thematic of the present work (correlation effects in strong-field ionization), shows  that, precisely at $R=5.0$\,a.u., this pattern of equidistant interference fringes is shifted and distorted,  and the nodal structure blurred as the photoelectron spectrum carries the signature of a multi-orbital ionization. The detailed analysis of these spectra, with a comparison with those associated with the SAE, or a non-correlated dynamics, is too long to be presented here, and exceeds the scope of the present paper. It will be presented in a separate contribution. It suffices to say that observation of this non-SAE signature in the channel-resolved photoelectron or LIED spectra in coincidence with the observation of a non-monotonous behaviour of the total ionization probability as a function of the field intensity would suffice to establish this electron-correlation driven ionization dynamics.


\section{Conclusions}
\label{sec:Conclusion}

We have set out to explore  possible manifestations of the limit of the  SAE  approximation in the description of the intense-field ionization of H$_2$. To this end, we used a model of the molecule in a finite function basis, as customarily done in Quantum Chemistry, specifically the $6\hbox{-}31$G$^{**}$ basis set. It corresponds to a time-dependent version of a quantum chemical full-CI representation with an active space of two-electron CSFs spanned by the most relevant molecular orbitals, the charge-resonance pair $\sigma_g$ and $\sigma_u$. The TDCI (with Feshbach partitioning) algorithm that we have access to\,\cite{nguyen-dang_multicomponent_2013, doi:10.1063/1.4904102} allows one to solve the many-electron TDSE, here with the possibility of tuning at will the electron correlation, through the introduction of an adiabatic switching-off of the two-electron interaction potential $V_{ee}$. The effect of switching off this interaction depends on the strength of electron correlation and this depends on $R$.  We have focused on  three values of $R$ typical of three regions of progressively increasing electron correlation. The equilibrium one $R=1.4$\,a.u, an elongated geometry $R=5.0$\,a.u, and a geometry at the dissociation limit $R=10.2$\,a.u. 

The observable we have addressed is the total ionization probability profile as a function of the field intensity. This profile follows a regular and nearly smooth increasing behavior, both at the equilibrium geometry and at the dissociative  limit. A sudden probability jump in the highest range of the field intensity, is observed however at $R=1.4$\,a.u. This sudden increase of the total ionization probability is interpreted as the onset of an over-the-barrier ionization regime. The most striking observation is however found in  an elongated, (but not dissociative), geometry, such as $R \simeq 5$\,a.u. There, the total ionization probability profile exhibits a non-monotonous behavior,  passing through a rise to a maximum  then a dip, denoting a partial quenching of the ionization, at some moderate intensity, the value of which depends on the field frequency. The value of the Keldysh parameter for ionization out of the ground-state then pertains to the intermediate regime \mbox{($\gamma\sim 0.5$)}, indicating a  possible competition between tunnel and multiphoton ionizations. We provide an interpretation of this non-monotonous variation of $P_{ion}(t_f)$ with $I$, referring to an interference mechanism among two overlapping resonances, corresponding to the autoionization  of a pair of dressed excited states, reached by a multi-photon REMPI process. Note that this interpretation is based on considerations of the correlation-dependent molecular energy spectrum, where the positions of the excited states with respect to the ionization threshold matter as well as the strong transition moments linking  them to the ground state.

\begin{acknowledgments}
Jean-Nicolas Vigneau is grateful to the French MESRI (French Ministry of Higher Education, Research and Innovation) for funding his PhD grant through a scholarship from EDOM (Ecole Doctorale Ondes et Mati\`ere, Universit\'e Paris-Saclay, France). JNV also acknowledges partial funding from the Choquette Family Foundation - Mobility Scholarship and the Paul-Antoine Gigu\`ere Scholarship. Numerical calculations conducted in Canada used HPC resources of the Compute Canada and Calcul Qu\'ebec Consortia (group CJT-923). This work was also performed using HPC resources from the ``M\'esocentre” computing center of CentraleSup\'elec and \'Ecole Normale Sup\'erieure Paris-Saclay supported by CNRS and R\'egion \^Ile-de-France. We finally acknowledge the use of the computing cluster MesoLum/GMPCS of the LUMAT research federation (FR 2764 at Centre National de la Recherche Scientifique). T.T.N.D. acknowledges partial funding of this research by the Natural Science and Research Council of Canada (NSERCC) through grant 05369-2015.
\end{acknowledgments}

\section*{Data Availability Statement}
The data that support the findings of this study are available from the corresponding author upon request.

\section*{Bibliography}

\nocite{*}
\bibliography{main}

\end{document}